# Destabilization of Alzheimer's Amyloid-β Protofibrils by Baicalein: Mechanistic Insights from All-atom Molecular Dynamics Simulations

Sadika Choudhury[1] and Ashok Kumar Dasmahapatra[1,2*]

[1]Department of Chemical Engineering, and [2]Centre for Nanotechnology, Indian Institute of Technology Guwahati, Guwahati – 781039, Assam.

## Abstract

Alzheimer's disease (AD) is a neurodegenerative disorder; it is the most common form of dementia and the fifth leading cause of death globally. Aggregation and deposition of neurotoxic Aβ fibrils in the neural tissues of the brain is a key hallmark in the pathogenesis of AD. Destabilisation studies of the amyloid-peptide by various natural molecules are of the utmost relevance due to their enormous potential as neuroprotective and therapeutic agents for AD. We performed molecular dynamics (MD) simulation on the U-shaped pentamers of amyloidogenic protofilament intermediates to investigate the destabilisation mechanism in the presence of Baicalein (BCL), a naturally occurring flavonoid. We found that the BCL molecule formed strong hydrophobic contacts with non-polar residues of the protofibril. Upon binding, it competed with the native hydrophobic contacts of the Aβ protein. BCL loosened the tight packing of the hydrophobic core of the protofibril by disrupting the inter-chain salt bridges and hydrogen bonds. The decrease in the structural stability of Aβ protofibrils was confirmed through the enhanced root mean square deviation (RMSD), radius of gyration and solvent accessible surface area (SASA), and reduced β-sheet content. PCA indicated that the presence of the BCL molecule intensified protofibril motions, particularly affecting residues in Chain A and B regions. Our findings propose that BCL would be a potent destabiliser of Aβ protofilament, and may be considered as a therapeutic agent in treating AD.

[*]Corresponding author: Phone: +91-361-258-2273; Fax: +91-361-258-2291; Email address: akdm@iitg.ac.in

# 1. Introduction

Alzheimer's disease (AD) is a neurodegenerative disorder defined by the progressive loss of cognitive, behavioural, and social abilities impairing a person's capacity to operate independently. It is the most common form of dementia and the fifth leading cause of death globally[1]. The risk of developing AD is closely related to old age, so it is estimated to increase by five million annually due to the ageing global population. Consequently, the prevalence of AD is expected to double every two decades, putting a significant burden on healthcare systems [2,3] Therefore, early diagnosis and preventive measures are necessary to slow down the disease's progression and reduce its incidence.

AD is characterised by the accumulation of hyperphosphorylated intraneuronal neurofibrillary tangles (NFTs) and extracellular Amyloid-β (Aβ) plaques[4]. Aβ peptide is derived by the sequential cleavage of the amyloid precursor protein (APP), an integral membrane protein by beta and gamma-secretase [5]. The γ-secretase enzyme generates two primary forms of peptides- Aβ40 and Aβ42, which consist of 40 and 42 amino acid residues, respectively. Aβ42 is the predominant form found in the diseased state and tends to aggregate more readily in solution, leading to increased neurotoxicity[6]. According to the amyloid cascade hypothesis (ACH), the discrepancy between Aβ generation and clearance causes dysfunction and, ultimately, neuronal death[7]. Other factors that are also considered to contribute to the pathogenesis of AD include cholinergic dysfunction[8] and oxidative stress[9]. As AD progresses, the loss of neurons in the cerebral cortex causes brain atrophy, and the connection between the temporal lobe and the hippocampus is severed[10]. These issues significantly impact a person's ability to remember, think, communicate, and make decisions.

The Aβ aggregation occurs most likely through three distinct pathways: specific on and off-pathways and a non-specific amorphous pathway [11]. The oligomerisation of Aβ forms aggregates such as amyloid fibrils, irregular β-aggregates, and amorphous unstructured aggregates. The on-pathway chemical process involves the polymerisation of highly ordered intermediates derived from monomeric aggregation-prone state (APS) β-sheet structures. The ordered intermediate prefibrillar aggregates ultimately form the fibril components of amyloid plaques. On the other hand, the off-pathway process involves converting irregular APS β-hairpin monomers into oligomeric aggregates, which do not form toxic fibrils. Furthermore, the other possible non-specific pathway involves misfolding and oligomerising disordered coil monomers into tangled amorphous and unstructured aggregates. As the off-pathway and non-specific mechanisms do not result in fibril formation, the constructed aggregates are initially non-toxic to cells[11]. Molecular medicine breakthroughs have put the amyloid-β pathway at the heart of AD pathophysiology. Diverting on-pathway Aβ aggregates into off-pathway irregular aggregates or amorphous agglomerates suppresses fibril formation and is a promising therapeutic strategy for AD.

One of the key methods to inhibit Aβ protein aggregation involves using molecules that can target intermediates formed in the aggregation pathway. Aβ fibrillogenesis has been targeted using antibodies[12], peptide inhibitors[13], non-peptidyl small molecules[14] and other natural compounds with anti-amyloidogenic properties[15,16]. Flavonoids and polyphenols commonly found in fruits and vegetables have been found to offer protective effects against AD[17]. They can interact with the aromatic rings and hydroxyl groups of the proteins preventing their aggregation. Furthermore, they can interact with the β-sheet structures of Aβ peptides[18]. Flavonoids like myricetin and morin exhibit comprehensive binding capabilities across various segments of the Aβ peptide, including β1, β2, and turn regions. Their binding efficacy is rooted in the presence of hydrogen bond donor and acceptor groups such as ether, carbonyl, and hydroxyl groups within their structure, leading to the disruption of the Aβ peptide's structure[19]. Morin was found to attach to fibril tips, impeding the binding of incoming peptides and reducing polymerization speed[20]. Wine-related polyphenols, including myricetin, morin, and quercetin, also demonstrate dose-dependent inhibition of fibrillar amyloid-beta formation and extension[21]. Similarly, Epigallocatechin-3-gallate (EGCG) from green tea was found to modulate molecular pathways, reducing Aβ accumulation in AD[22]. Resveratrol, another flavonoid with antioxidant properties, ameliorates cognitive function in neurodegenerative diseases and protects against Aβ-induced neuronal disruption, as demonstrated by Karuppagounder et al.[23] and Wang et al. [24]. Experimental and theoretical literature underscores the multifaceted impact of flavonoids, inhibiting both early and late stages of Aβ aggregation.

Computer simulations by exploring different time and length scales can complement experiments and provide insights into the destabilisation process. Furthermore, they also validate in vitro and in vivo results. In this study, we used MD simulation to investigate the physical and structural changes in Aβ protofibril upon interacting with Baicalein (BCL). Baicalein(5, 6, 7-trihydroxy flavone) is a naturally occurring flavonoid found in Chinese medicinal plants belonging to the Scutellaria genus and has also been reported in Oroxylum indicum (Indian Trumpet flower)[25] and Thyme[26]. This flavone has recently attracted interest for its capacity to halt the progression of AD and other neurodegenerative diseases. BCL was found helpful in alleviating cognitive impairment related to AD[27]. BCL was also shown to reduce oxidative stress, inflammation, and the formation of certain amyloid proteins[28]. Additionally, it has been found to have anti-aggregating and anti-amyloidogenic properties for various proteins, including Aβ-peptide[29], α-synuclein[27,30], and lysozyme[31]. It has been reported that BCL inhibit the fibrillation process and disassemble existing fibrils of wild-type α-synuclein in vitro[32]. Additionally, BCL stabilise α-synuclein demonstrates efficiency in impeding the fibrillation of untreated alpha-synuclein[33]. Investigations into the influence of baicalein on mutant α-synuclein-induced aggregation have revealed its effectiveness in mitigating the aggregation of E46K mutant α-synuclein, both in in vitro experiments and cell culture systems, leading to a reduction in associated cell death[34]. In a recent study, BCL was tested against SOD1 fibrillation[35], which yielded notable results—a significant inhibition of fibrillation and a reduction in the fibril load were observed. Notably, the flavonoid interacted with oligomers and prefibrillar species, ultimately leading to the destabilization of SOD1 fibrils. This interaction subsequently increased the concentration of monomers within the soluble fraction, thereby reducing toxicity. Therefore, these findings motivated the utilization of BCL as a destabilizing agent within the context of amyloid β fibril in our present study.

We have characterised the binding site and mode of the BCL molecule with the Aβ protofibril. The BCL molecule formed strong hydrophobic contacts with non-polar residues of the protofibril. Upon binding, it competed with the native hydrophobic contacts of the protein and brought conformation changes in the protofibril by disrupting the salt bridges and hydrogen bonds. Furthermore, metrics such as RMSD, $R_g$, and SASA exhibited a higher value in the presence of the BCL molecule, indicating a decrease in protofibril's stability. According to our findings, the BCL molecule can disrupt the protofibril and hence can be regarded as a possible therapeutic agent for treating AD.

## 2. Methods

### 2.1 Aβ42 protofibril model

A three-dimensional model of Aβ42(PDB ID: 2BEG), was taken for our current study [36]. The PDB database for 2BEG consists of all the ten models obtained by the solution NMR. However, the first model is typically used for simulations as it satisfies most structural constraints. The pentameric structure is a repeating unit of the fully-formed Aβ fibril and is best described as a protofibril[37]. It is a popular model used for several other studies as an intermediate conformer in the fibril system[38,39]. The protofilament comprises five chains A, B, C, D and E, as shown in Fig. 1A. The core region of the protofibril is formed by residues 17-42 but lacks the first sixteen N-terminus residues. The first sixteen residues are disordered however contribute significantly less to the fibril's overall stability. Hence, the observation made using this model can be extended to the full-length fibril[40]. The U-shaped motif consists of two in-register β strands, β1 spanning from residue 18-26 and β2 from 31-42. The strands are connected by a bend region comprising residues 27-30.

## 2.2 Ligand Selection and Molecular Docking

We used a polyphenolic flavonoid, BCL, as a destabiliser, depicted in Fig. 1B. The initial structure of our ligand was obtained from PubChem[41], which was then used in docking. Molecular docking studies were performed using CB-Dock2[42], a blind docking web server. It is the updated version of CB-Dock[43]. Since its initial release, the CB-Dock web server has received over 200 task submissions per day from around the world. Numerous researchers have used it for docking studies, and in particular, it was widely used in investigating COVID-19 therapeutic agents[44,45]. Without any information about the binding site, CB-Dock uses a novel curvature-based cavity detection approach, CurPocket[46], to predict the binding site of a given protein and performs docking using Autodock Vina[47]. Ranking of the binding modes is done according to the Vina score. CB-Dock2 combines CB-Dock with a template-based docking engine, thereby improving binding site identification and pose prediction. In one of the benchmark tests, CB-Dock2 showed a 16% improvement in docking success rate compared to CB-Dock. We chose the docked complex in its best binding pose with the highest Vina score out of the top five poses generated by CB-Dock2. Before performing the MD simulations, it was submitted to Swiss-Param[48] to generate the parameters files required for GROMACS.

## 2.3 Molecular dynamics simulations

The best-docked complex was then subjected to MD simulation using GROMACS v5.1.1[49]. Each system was placed in a 6.8 nm cubic box and solvated. The water molecules were represented by the TIP3P model. Five sodium ions were added to neutralise the system. Periodic boundary condition (PBC) was applied in all the directions. Then energy minimisation using the steepest descent algorithm was done to reduce steric clashes and ensure an optimised starting structure. Short-range non-bonded interactions were truncated at 1.2 nm. LINCS[50] was used as a constraint algorithm for bond length. The Particle Mesh Ewald (PME) method[51] was used for long-range electrostatic interactions, with Fourier grid spacing of 0.16 nm. Before the production run, the minimised systems were equilibrated with position restraints applied to heavy atoms. A 100 ps NVT equilibration was followed by a 400 ps equilibration under NPT ensemble conditions to stabilise the temperature and pressure at 300 K and 1 bar, respectively. Velocity rescaling[52] and Parinello Rahman barostat[53,54] were employed for temperature and pressure coupling, respectively. The equilibrated systems obtained were then submitted for a 300 ns MD run after removing the position restraints. The MD simulation has been carried out using CHARMM36m(Chemistry at HARvard Macromolecular Mechanics) all-atom force field[55]. It was used in recent studies on Aβ42[56], and some studies even recommended using it for amyloid fibril simulation[57].

## 2.4 Analysis Details

The protein-ligand interactions were studied using the ProteinPlus [58]webservers and LigPlot+ software[59]. The inbuilt tools in GROMACS package, VMD[60] and UCSF-Chimera[61] were used to measure the various parameters for our analysis. The global structural stability of the protofibril was assessed by $C_\alpha$-RMSD using g_rms. The radius of gyration ($R_g$) and solvent-accessible surface area (SASA) values were calculated using g_gyrate and g_sasa. For the secondary structure analysis, the DSSP tool was used. The Molecular mechanics Poisson−Boltzmann surface area (MM/PBSA) method [62]was employed to calculate the interaction free energy of the BCL molecule with the amyloid fibril. The package was used within GROMACS by the g_mmpbsa[63] tool. Data extracted at every 100ps from the last 10ns of the trajectory was considered for our MM/PBSA analysis. The solute and solvent dielectric constants were taken as 4 and 80, respectively. The MM/PBSA method uses the following expression to calculate the binding free energy(BFE)[64]:

$$\Delta G^{bind} = \Delta E_{MM} + \Delta G^{psolv} + \Delta G^{npsolv} - T\Delta S$$

In this equation, $\Delta E_{MM}$ stands for the molecular mechanics contribution to the BFE in vacuum, $\Delta G^{psolv}$ and $\Delta G^{npsolv}$ are the electrostatic and non-electrostatic contributions to the solvation energy. They were calculated using Poisson–Boltzmann equation and the SASA model, respectively. The high computational load constrains the calculation of the entropy term. Furthermore, the relative change in the entropy ($\Delta S$) contribution towards the BFE calculation is insignificant and hence can be excluded[56]. The individual residue contributions were also calculated using g_mmpbsa. GROMACS modules g_covar and g_anaeig were used for Principal Component Analysis (PCA). The interchain distances were plotted to determine the residue-residue contacts between adjacent chains. The structural strength of the protofibril was determined by measuring the change in the number of hydrogen bonds across the trajectory. The inter-chain salt bridges and hydrophobic contacts for different residue pairs of all the neighbouring chains were evaluated using the GROMACS g_mindist module. Each system was simulated three times to check for the reproducibility of results. The data thus represented are average over three sets.

## 3. Results and discussion

### 3.1 Molecular docking, evaluating Aβ-BCL binding mode

The BCL molecule was first docked to the three-dimensional structure of Aβ42 oligomer (PDB ID: 2BEG), using CB-Dock2, a blind docking software. Without prior knowledge of the target pockets, blind docking protocol scans the entire surface of the receptor protein for possible binding sites. We chose the highest-ranked pose with a binding energy of -6.9 kcal/mol. The docked complex was visualised using UCSF-Chimera, as shown in Fig. 2. The high interaction energy signifies that BCL is an excellent binder to the Aβ protofibril, and can potentially bring conformational changes in the protein.

The BCL molecule binds to the outer U-shaped cavity of the protofibril, formed by the β1 and β2 strands. In the Aβ42 model considered in our study, the L17 residue points outward while F19 is directed inwards, creating a large pocket surrounded by the terminal residues: L17, V18, F19, G38, V39, and V40. The other protein cavities are buried deep inside the fibril. Thus, the larger volume of the outer cavity and its proximity to the solution environment make it a more favourable binding site.

The analysis of the docked structure reveals that besides forming hydrophobic interaction, the BCL molecule also forms a hydrogen bond with the V39 residue of chain E and pi-stacking with the F19 residue of chain D, as shown in the protein-ligand interaction diagram (Fig. 2). The residues in contact with the ligand were also tabulated (Table 1) and depicted in Fig. S1. The interaction of the BCL molecule with the hydrophobic residues of protofilament enables its access to the hydrophobic cavity. The findings were consistent with previous studies where compounds docked in the outermost cavity were found to destabilise the Aβ oligomer[65,66].

## 3.2 Validation of computational data with experimental results

Before assessing the impact of the BCL molecule on the Aβ42 protofibril, we validated the conformational ensemble generated through molecular dynamics (MD) simulations against experimental structures. This involved calculating NMR chemical shift values for $C_\alpha$ and $C_\beta$ atoms using SHIFTX2[67], which were then compared with experimental data[68]. The average computational $C_\alpha$ and $C_\beta$ chemical shifts ($\delta_{sim}$) showed a robust correlation of 0.94 and 0.99, respectively, with the experimental chemical shifts ($\delta_{exp}$), as depicted in Fig. S4 (Supplementary Information). The comparison demonstrated strong agreement between experimental and computational data for the Aβ42 protofibril, consistent with previously reported results[69], affirming the reliability of our MD simulation data. Additionally, three-bond J–coupling ($^3J_{NH-H\alpha}$) constants were also analysed. These constants reflect the three-bond coupling interaction between HN and Hα protons, providing further insights into peptide conformations. The three-bond J-coupling constants, shown in Fig. S5 (Supplementary Information), were derived via the Karplus equation[70] using the Vuister and Bax parameters[71]. The J coupling constants of the Aβ42 protofibril show significant agreement between experimental and simulated outcomes.

### 3.3 MD simulation of Aβ protofilament in the presence of BCL

The simulations were carried out for 300 ns for all the systems following the protocols discussed in the method section. The final configurational snapshots of the systems at 300 ns were taken. We can observe from Fig. 3A that the protofibril in the control system was reasonably stable and showed only slight twisting of the peptide chains at the edges, and a marginal outward movement of chain A. The twists of the β strands increase protofibril's stability further by increasing the side chain packing[72]. However, the peptide chains has been highly destabilised by the introduction of BCL molecule, as shown in Fig. 3B. The disorganised structure of the protofibril in the Aβ-BCL systems prompted us to study the disruptive effect of the ligand further. Hence, other parameters were evaluated, as discussed below.

#### 3.3.1 Destabilisation of the Aβ oligomer

In the presence of the BCL molecule, the Aβ oligomer was destabilised. The degree of destabilisation was measured by using various global stability parameters, such as, $C_\alpha$-RMSD, $R_g$ and SASA. The average values of these parameters across the 300 ns trajectory were plotted for both the Aβ-BCL and Aβ-Water systems and were compared quantitatively.

The $C_α$-RMSD of the protein was used to assess the structural stability of the protofibrils. It was found that the average RMSD value was increased to 1.17 ± 0.13 nm in the presence of the BCL molecule as compared to the 0.92 ±0.09 nm of the control (viz., Aβ-Water) system. The same can also be observed from the average $C_α$-RMSD plot for the protein (Fig. 4A), where the curve for the Aβ-BCL is shifted upward towards a higher RMSD, unlike the BCL-free system. A similar trend in the RMSD plot was observed in previous destabilisation studies of the oligomer[73,74]. The twisting of chains A and E has contributed the most to the RMSD of the control systems. This can be attributed to the limited contact that peripheral chains have with other chains, leading to a tendency to deform in order to maximise contact[75]. However, the overall U-shaped motif formed by the β1 and β2 strands was preserved.

The RMSD values from the β1, β2, and turn regions were calculated to expound more information about the most affected areas of the protofibril. It was found that all the regions showed an increased RMSD in the presence of the BCL molecule. The average RMSD for β1, β2 and turn region has been increased from 0.66 ± 0.10 nm, 1.07 ± 0.08 nm and 0.96 ± 0.16 nm in the Aβ-Water system to 1.09±0.20 nm, 1.18 ±0.09 nm and 1.09 ±0.13 nm in the Aβ-BCL system, respectively. The difference in the RMSD value is more pronounced in the β1 region, as can be observed from Fig. 4B. The average RMSD values of these regions were plotted with the associated standard deviation represented as error bars. The increased RMSD suggests that the BCL molecule caused significant conformational changes in the protofibril.

The Radius of gyration ($R_g$) measures the compactness of a protein structure. Therefore, we measured the $R_g$ values to evaluate whether the BCL molecule could increase the mobility of the chains, and hence decreasing the rigidity of the fibril structure Our results align well with the RMSD analysis. The average $R_g$ value of the protein was found to be 1.47 ± 0.02 nm and 1.52 ± 0.02 nm for Aβ-Water and Aβ-BCL systems, respectively (Fig. 5A). A comparable increase in $R_g$ values was also observed in a previous study focusing on the destabilization of Aβ protofibrils using Arginine-containing short peptides[76].

We have measured the solvent-accessible surface area (SASA) of the protofibril (Fig. 5B) to estimate how tightly the chains are packed. The non-polar residues are packed together in the hydrophobic core during protein folding, and they are essential for maintaining protein structure and stability. The average SASA value of the fibril has been increased significantly from 81.09±2.19 nm$^2$ to 86.61±2.21 nm$^2$ in the presence of the BCL, indicating loosening of the fibrillar structure. Thus, increasing SASA suggests that the BCL molecule is capable in disrupting the hydrophobic packing of the protofibril by increasing its exposure to water. The increase in both the average $R_g$ and SASA values in the Aβ-BCL system indicates that the perturbation has been escalated in the presence of BCL, proving its potency as a destabiliser.

### 3.3.2 Changes in the secondary structure

Aβ monomers with initial β-sheet structures self-assemble into oligomers and aggregate into toxic fibril via a primary nucleation mechanism [77]. The formation of β-sheet features is a critical early step in Aβ amyloidogenesis and neurotoxicity[78]. Hence, evaluating the change in secondary structure of the protofibril would be valuable in comprehending the anti-aggregation effect of the flavonoid. We assessed the impact of the BCL molecule on the β-sheet structure content of Aβ peptide in the pentameric protofibril state using the DSSP (dictionary of secondary structure of proteins) method.

Fig. 6A illustrates that the Aβ protein had a high β-sheet content in the water system. Upon interaction with the BCL molecule, there was a noticeable increase in coil structures at the expense of β-sheet content over the 300 ns trajectory, as shown in Fig. 6B.
It is important to note that converting the β-sheets into coil or α-helices is a crucial step in impeding primary nucleation, and preventing the amyloid fibrillation process. Although the Aβ-BCL system preserves the β-strand to a certain extent, there were still noticeable alterations compared to the Aβ-Water system. There was particularly an increase in coil content and a decrease in β-sheet content, as shown in Table 2. The β-sheet content has been decreased from 37% to 30%, while the coil content has been increased from 39% to 46% in the system with the BCL molecule. The present findings are consistent with previous studies that have reported a reduction of β-sheet content of Aβ fibril resulting in its destabilisation upon exposure to D744, a fluorinated derivative of curcumin[79], and norepinephrine[80].

### 3.3.3 Binding mode of BCL molecule

The MM-PBSA method was integrated with MD simulation to investigate the binding affinity of the BCL molecule with the protofibril chain and to calculate the energetic contribution of various interactions. For the MM-PBSA analysis, the last 10 ns of the trajectory was considered with $\Delta t$ = 10 ps. It was observed that both the non-bonded van der Waal (vdW) and electrostatic interactions favoured ligand binding (Table 3). However, the vdW interaction was almost ~10-fold more dominant than the electrostatic interaction. The contribution of the vdW energy (-146.64 ± 8.52 kJ/mol) dominates over all the other energy terms, contributing the most towards the negative $\Delta G_{binding}$ of 76.59 ± 10.23 kJ/mol. The non-polar solvation energy (-15.75 ± 0.83 kJ/mol), is also favourable for the formation of the complex. However, the same cannot be said for the polar component of solvation energy.

To gain further insights into the major amino acid residue involved in binding, we have calculated the individual contribution of the residues across the entire length of the protofibril. It can be noted from Fig. 7A that chains A and B contributes the most towards the overall interaction energy. Subsequently, we have calculated the individual contribution of residues from these chains. As illustrated in Fig. 7B and C, our analysis reveals the key involvement of the non-polar hydrophobic residues - F19, A21, V24, and I32 of chains A and B.

The ability of BCL to interact with the hydrophobic residues of the core contributes to higher van der Waals interaction energy, as calculated by the MM-PBSA analysis. Supplementary videos S1 and S2 showcase the configurational changes in the protofibril structure throughout the 300 ns trajectory, both in Aβ-water and Aβ-BCL systems. The control system demonstrates slight twisting in the protofibril, whereas the presence of the BCL molecule significantly destabilizes the protofibril structure. The contacts established by the BCL molecule with the protofibril residues at the 300 ns mark are illustrated in Fig. S2 of the Supplementary Information. The aromatic ring of Phenylalanine and the side chain of Isoleucine residues play a vital role in ligand binding. Furthermore, the participation of the F19 residue in the protein-ligand interaction has been elucidated[81] in Fig. S3. It was observed that initially, F19 formed π-stacking and hydrophobic interactions; however, towards the end, it predominantly exhibited hydrophobic interactions. This observation suggests that although other interactions play a role in the protein-ligand interaction, hydrophobic interactions ultimately drive the destabilization process. The residue-wise energy contribution was also calculated for chains C, D and E, as depicted in Fig. S7 (Supplementary Information). Compared to chains A and B, these chains demonstrate a lower degree of interaction with the BCL molecule. The hydrophobic residues I32, L34, and V36 present in chains C, D and E contribute to their interaction with the BCL molecule.

The average BFE was also computed for the adjacent chains in both the presence and absence of the BCL molecule. The inter-chain BFE, detailed in Table S3 of the Supplementary Information, indicated that in the presence of BCL molecule, the chains exhibited a higher BFE. This observation substantiates that, in the presence of the ligand, the adjacent chains of the protofibril displayed a diminished binding affinity, resulting in an increased distance between the chains.

We then carried out contact map analysis to qualitatively assess the effect of the ligand on the residue-residue contact between adjacent chains of the oligomer (Fig. 8). The contact map analysis provides the smallest distance between the $C_\alpha$ atoms of two residue pairs in Aβ42 protofibrils. The panel diagonals display the most substantial binding, representing homologous interactions between two neighbouring chains in close contacts, such as N- and C-termini from one chain with an equivalent portion from the neighbouring chain. The off-diagonals display the lateral non-homologous interaction, i.e. N-terminal interactions of one chain with the C-terminus of the other chain or vice-versa. The analysis was performed with the gmx mdmat module of GROMACS. The matrices were created using a minimum distance of 0–1.5 nm that separates the residue pairs. We have observed a drastic decrease in contact of the Aβ-BCL system compared to Aβ-water. The inter-chain contacts between Chain A, B and C were fairly compromised in the presence of the BCL molecule. The off-diagonals non-homologous interactions were mostly affected, as can be inferred from the colour range of the matrix shown in Fig. 8C and D. The contacts remained relatively sustained in Aβ-Water systems, as shown in Fig. 8A and B. The same trend in the inter-chain contacts was also observed between chains C, D and E, as depicted in Fig. S8 (Supplementary Information).

The compact oligomeric aggregates are produced by the strong inter-chain electrostatic and van der Waals interaction energies. The increased interchain distance, corroborated by the reduced inter-chain binding affinity, shows that the parent fibrillar structure in the Aβ-BCL systems moved from the highly ordered configuration towards a more destabilised form, with the chains further away. The MM-PBSA results and the decreased contact between the adjacent chains corroborate well, establishing BCL molecule as a promising destabilising agent of Aβ42.

## 3.4 Principal Component Analysis

Principal Component Analysis (PCA) serves as a multivariate statistical method employed to systematically reduce the necessary dimensions for describing protein dynamics. PCA operates as a linear transformation, extracting crucial elements in the data using a covariance matrix derived from atomic coordinates. PCA reads the MD simulation trajectory and extracts the dominant modes in the motion of the molecule. This reduction is achieved through a decomposition process that sifts observed motions from the most substantial to the smallest spatial scales. The analysis of principal components (PCs) can provide valuable insights into the predominant modes of motion exhibited by the protofibril in isolation and in complex with the BCL molecule. To illustrate alterations in motion induced in the protofibril by the interaction with BCL, PC analysis was conducted on the MD trajectories during the last 10 ns of simulation for Aβ42. Fig. 9A exhibits plots of the eigenvalues against eigenvector indexes obtained from the diagonalization of the covariance matrix, where the eigenvalues reflect the intensity of the movements, and the eigenvector indicates movement directions. We observe that the amplitude of the first few eigenvalues decreased to reveal several constrained, more localized fluctuations. The analysis shows that the first, second, and third PCs account for 51.3%, 20.2%, and 9.35% in the Aβ-BCL system and 31%, 27.25%, and 12.7% in the Aβ-water system. The elevated eigenvalues observed in the Aβ-BCL system, in comparison to Aβ-water, suggest an increased magnitude of motion in the protofibril in the presence of the BCL molecule. To assess the impact of the BCL molecule on the motion represented by PC1, corresponding to the direction of maximum variance, the displacements of PC1 were calculated for both systems. As depicted in Fig. 9B, the results suggest that the BCL molecule influences the motions of the protofibril. In the control system, the motion is attributed to twisting, while in the presence of the BCL molecule, this motion is exacerbated. Notably, residues in Chain A and B regions exhibit elevated values, aligning with our MM-PBSA studies, where we observed that residues in these chains contribute significantly to the overall interaction.

### 3.5 Effect of BCL on various bonds of Aβ fibril

Intra- and inter-molecular bonds play a crucial role in maintaining the stability of protein structures. These bonds include hydrogen bonds, covalent bonds, salt bridges, and hydrophobic contacts. Understanding the changes in these molecular interactions in the presence of an external molecule would provide insights into its impact on protein stability. We have conducted a detailed analysis to elucidate the effect of BCL molecule on the critical molecular interactions that stabilise the Aβ42 (2BEG) structure.

### 3.5.1 Hydrogen Bond

Hydrogen bonds (H-bonds) are crucial in forming secondary structures and higher-order protein aggregates. They contribute significantly to structural integrity of protein. In Aβ fibrils, there is an extensive intra- and inter-peptide H-bonding network. Several experimental and theoretical investigations have demonstrated the importance of these H-bonds in stabilising Aβ fibrils [72,82].

We have calculated the average number of H-bond over 300 ns trajectory for the entire protein. The number of H-bond has been reduced from $59 \pm 5$ to $52 \pm 5$. The decreasing trend of the average number of H-bond can also be observed in Fig. 9. The average inter-chain H bond has also been tabulated in Table S4 (Supplementary Information). There is an observable decrease in H bonds between chains A-B and B-C in Aβ-BCL systems. The number decreased from $11\pm2$ to $9\pm1$ between chains A-B and from $15\pm1$ to $10\pm1$ between chains B-C. The reduction in H-bond indicates the increased instability of the protofibril. These observations are in close agreement with the destabilisation of Aβ fibril by a proline-rich β-sheet breaker peptide [83].

### 3.5.2 Salt Bridge

A salt bridge is formed between two oppositely charged amino acid residues, typically between an acidic residue such as glutamate or aspartate and a basic residue such as arginine, lysine or histidine. The interaction between these residues results in the formation of an electrostatic bond. Salt bridges are essential structural features of proteins, and play a critical role in stabilising the three-dimensional structures of the protein [84]. In the protofibril structure chosen for the current study, the salt bridges formed between D23 and K28 residues are more prominent than those formed between E22 and K28 residues [36]. We report only those significant interchain salt bridges (viz., D23 and K28), by measuring the average distance between D23 and K28.

It can be observed from Fig. 10 that the average D23-K28 distances for all the four sets of neighbouring chains increased across the trajectory for the Aβ-BCL system. In contrast, the distance for the Aβ-Water system does not show much changes, rather remains relatively stable. Although, there was an increase in the average D23-K28 distance for chain A even in the control system, it was less prominent than in the Aβ-BCL system. The increase in average distance for chain A in the control system was because of the inherent configuration of the K28 residue. The K28 residue of chain A points away from the D23, resulting in a weaker electrostatic bond. This weaker electrostatic bond explains the slight outward movement of chain A, as has been observed in Fig. 3A. However, the Aβ-BCL system shows a notable increase in the average distance between the D23 and K28 residues of all other adjacent chain pairs compared to the control systems. The increased distance signifies the weakening and disruption of these inter-chain salt bridges, which are crucial for the overall stability of the protofibril.

### 3.5.3 Hydrophobic contacts

Besides electrostatic interactions, hydrophobic interactions also play a crucial role in protein folding and stability. The hydrophobic residues buried inside protein cores enhance protein stability. Studies on Aβ fibrils have reported the vital role of hydrophobic contacts in the structural organisation and strengthening of the fibrils [85]. The hydrophobic contact pairs A21-V36, L34-V36, and F19-G38 have been identified as significant contributors towards the stability of the current Aβ pentamer structure [36].

As discussed earlier, the MM-PBSA analysis revealed that the BCL molecule firmly binds to the F19 and A21 residues of chain A and B. We assumed that the binding of the BCL molecule to F19 and A21 residues may have disrupted the crucial hydrophobic bonds associated with these residues. Therefore, we calculate the average distances between the inter-chain A21-V36 and F19-G38 residues to study the disruption of hydrophobic contacts, as the inter-chain residue distance measures the compactness of the protofibril structure.

As observed from Fig. 11, all the average inter-chain A21-V36 and F19-G38 distances has increased significantly in the presence of the BCL molecule. The only deviation was observed in chain A, where the distance between the A21-V36 residues was slightly higher in the Aβ-Water system. As explained earlier, the orientation of the K28 residue of chain A in the native protofibril model results in a weaker electrostatic bond between chains A and B, making chain A more susceptible to fluctuations. Therefore, this weaker electrostatic bond between chains A and B, explains the observed increase in the average A21-V36 distance, even for the Aβ-Water system. Hence, by binding to the hydrophobic residues, the BCL molecule has successfully destabilised the protofibril. In-silico study of the wgx-50 molecule has yielded similar results, demonstrating that disrupting these crucial hydrophobic bonds can destabilise the Aβ fibrils[38].

## 4. Conclusions

In addition to reducing neurotoxicity concerns, destabilisation of the oligomers can prevent the development of higher-order toxic aggregates. Therefore, inhibiting Aβ aggregation may be considered as a promising potential therapeutic strategy for AD. In this study, we employed all-atom MD simulations to examine the interaction of the BCL molecule with Alzheimer's amyloid β fibril. We compared our MD simulation results to experimental data by analysing NMR chemical shifts for $C_\alpha$ and $C_\beta$ atoms before studying the effect of the BCL molecule on Aβ42 protofibrils. It could be observed from the final snapshot of the Aβ-BCL system that the oligomer deformed in the presence of the flavonoid. The destabilising potential of the BCL molecule was investigated, and the binding sites and modes were characterised. Our study revealed that the hydrophobic interactions were the prominent driving force for the binding of the BCL molecule. The BCL molecule binds to the hydrophobic residues of the outer protofibril cavity. Upon binding, it competed with the native hydrophobic contacts of the protofibril. An increased inter-chain A21-V36 and F19-G38 distances demonstrate that the BCL molecule impeded the protofibril's ability to form hydrophobic contacts. Furthermore, the structural instability of the protofibril in the presence of BCL was confirmed via the increased values of parameters such as RMSD, $R_g$ and SASA. Additionally, the comparison of Principal Component (PC) 1 displacement for both systems indicated the influence of the BCL molecule on protofibril motion, notably evident in the elevated values of residues within the Chain A and B regions of the Aβ-BCL system. The BCL molecule brought conformation changes by disrupting the significant stabilising interactions of the protofibril. The inter-chain salt bridges and hydrogen bonding networks were disturbed, increasing coil content at the expense of β-sheets. The protofibrils loosened due to the weakening of the interatomic packing. The insights from our study showed that the BCL molecule could destabilise the protofibril and hence can be considered a potential therapeutic agent for the treatment of AD.

## Supporting materials

The specific contacts between the BCL molecule and the protofibril at the docking stage and after 300 ns of simulation have been depicted. Interactions at 0 ns and 290 ns have been shown to study interactions. The correlation between the predicted and experimental NMR chemical shifts and $^3 J_{HN-H\alpha}$ coupling constants has been plotted. Conformational snapshots at different timescales have been included. The $C_\alpha$-RMSD and $R_g$ values for all the runs have been depicted. The contribution of individual residues from chains C, D, and E to the total binding free energy has been plotted using bar graphs. The inter-chain distance matrix between chains C-D and D-E has been illustrated for both the Aβ-water and Aβ-BCL systems. The average inter-chain binding free energy has been tabulated. The average number of interchain hydrogen bonds has been tabulated for the systems. Additionally, the visual recording of the configurational changes of the protofibril in water and in the presence of the BCL molecule across the 300 ns trajectory has been presented. Videos VS1 and VS2 show the structural changes over 300 ns for the Aβ-water and Aβ-BCL systems, respectively.

## Author Information

Corresponding Author *Phone: +91-361-2582273. Fax: +91-361-2582291

E-mail: akdm@iitg.ac.in

ORCID Ashok Kumar Dasmahapatra: 0000-0002-0082-4881

The authors declare no competing financial interest.

## Disclosure Statement

No potential conflict of interest was reported by the authors.

# Acknowledgement

Computational facilities supported by the Department of Science and Technology (SR/FST/ETII-028/2010) and the HPC parallel computational facility, PARAM-ISHAN at IIT Guwahati, were used for the work reported in this paper and are highly acknowledged.

**Table captions:**

**Table 1: Aβ-BCL docking results.**

**Table 2: Percentage of different secondary structure contents of Aβ-Water, and Aβ-BCL systems during 300 ns simulation.**

**Table 3: Binding free energy between Aβ protofibril and BCL molecule.**

**Figure captions:**

Fig. 1: (A) Model of Aβ42 used in this study (PDB ID: 2BEG) (B) Initial structure of BCL molecule.

Fig. 2: Docked complex (Aβ –BCL), and protein-ligand interaction diagram.

Fig. 3: Conformational snapshot of (A) Protofibril in Aβ-Water, and (B) Protofibril and ligand in Aβ-BCL systems at 300 ns.

Fig. 4: $C_α$-RMSD values across the 300 ns trajectory of the (A) Protofibril, and the (B) average $C_α$-RMSD values β1, β2, and Turn region in control and Aβ –BCL system.

Fig. 5: (A) Radius of gyration ($R_g$), and the (B) SASA values of the protofibril in control and Aβ-BCL systems over 300 ns.

Fig. 6: DSSP secondary structure for (A) Aβ –Water, and (B) Aβ-BCL system over 300 ns.

Fig. 7: Individual residue contribution of (A) Entire fibril length, (B) Chain A, and (C) Chain B to the binding free energy of the protofibril and BCL molecule.

Fig. 8: Inter-chain distance matrix for (A) Chain A-B, (B) Chain B-C of the control and (C) Chain A-B, and (D) Chain B-C of the Aβ-BCL systems.

Fig. 9: (A) The eigenvalues plotted against the corresponding eigenvector indices, and (B) Displacements of the components of the Aβ-Water and Aβ-BCL for the first eigenvector.

Fig. 10: Change in the number of hydrogen bonds for Aβ-Water and Aβ-BCL systems over 300 ns simulation.

Fig. 11: The average inter-chain D23-K28 distances of the Aβ-Water and Aβ-BCL systems over 300 ns for (A) Chain A-B, (B) Chain B-C, (C) Chain C-D, and (D) Chain D-E.

**Fig. 12: Average inter-chain distances between (A) A21-V36, and (B) F19-G38 residues.**

**Table 1 : Aβ-BCL docking results.**

| Docked Complex | Vina Score (kcal/mol) | Residue in Contact |
|---|---|---|
| **Aβ-BCL** | -6.9 | **Chain B:** PHE 19<br>**Chain C:** LEU 17, VAL 18, PHE 19<br>**Chain D:** LEU 17, VAL 18, PHE 19, VAL 39, VAL 40<br>**Chain E:** LEU 17, VAL 18, PHE 19, GLY 37, GLY 38, VAL 39, VAL 40 |

**Table 2: Percentage of different secondary structure contents of Aβ-Water, and Aβ-BCL systems during 300 ns simulation.**

| System | % β-Sheet | % Coil | % Bend/Turn/Helix |
|---|---|---|---|
| **Aβ - Water** | 37 | 39 | 24 |
| **Aβ - BCL** | 30 | 46 | 24 |

**Table 3 : Binding free energy between Aβ protofibril and BCL molecule.**

| Energy terms | Aβ-BCL (kJ/mol) |
|---|---|
| $\Delta E_{vdW}$ | -146.64±8.52 |
| $\Delta E_{elec}$ | -15.49±5.90 |
| $\Delta G_{ps}$ | +101.30±8.56 |
| $\Delta G_{nps}$ | -15.75±0.83 |
| $\Delta G_{binding}$ | -76.59±10.23 |

(A)

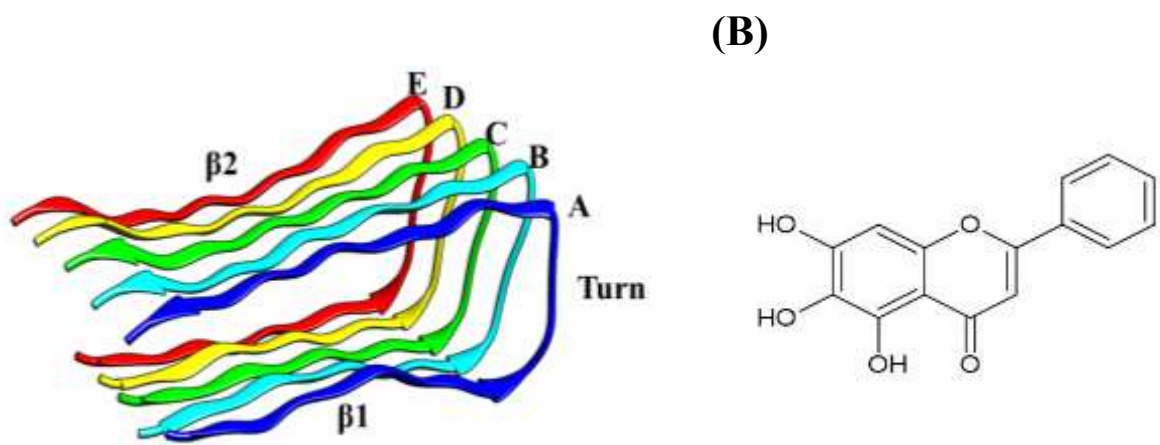

**Fig. 1: (A) Model of Aβ42 used in this study (PDB ID: 2BEG) (B) Initial structure of BCL molecule.**

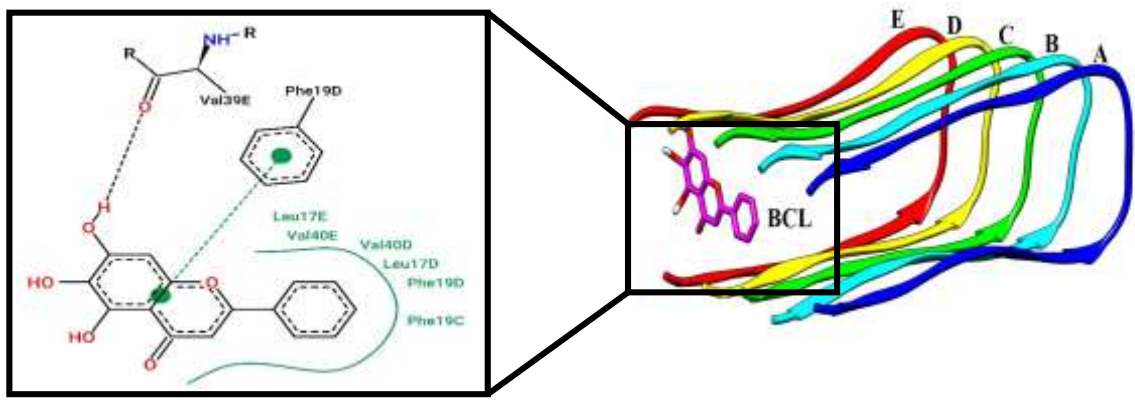

**Fig. 2: Docked complex (Aβ –BCL), and protein-ligand interaction diagram.**

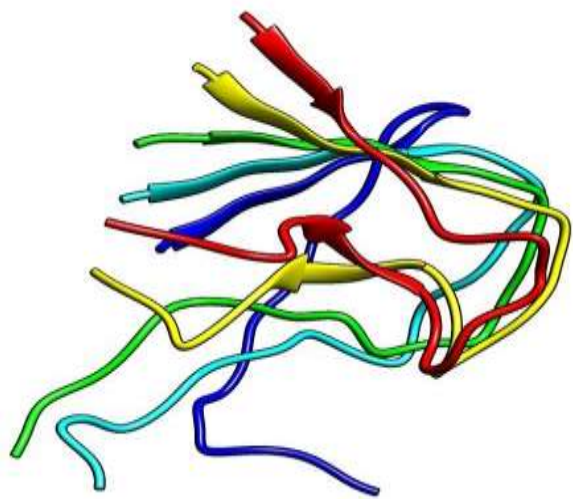 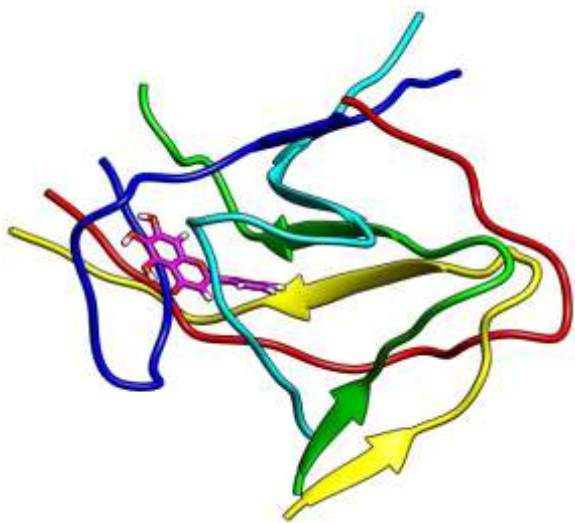

**Fig. 3:** Conformational snapshot of (A) Protofibril in Aβ-Water, and (B) Protofibril and ligand in Aβ-BCL systems at 300 ns.

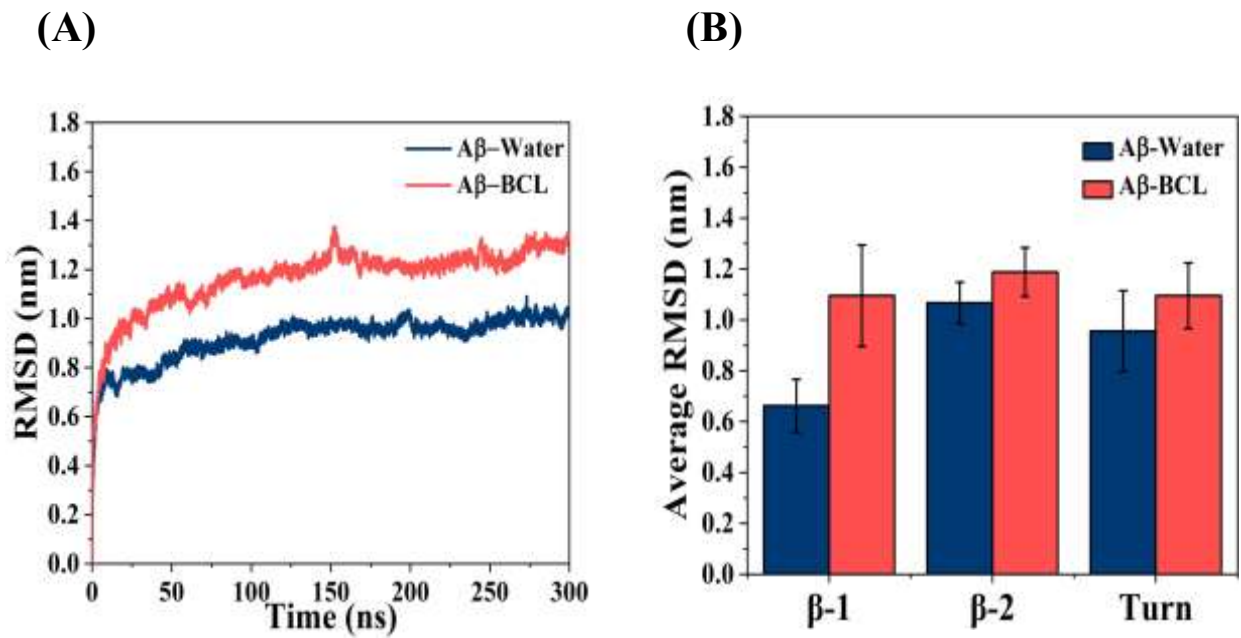

**Fig. 4:** $C_\alpha$-RMSD values across the 300 ns trajectory of the (A) Protofibril, and the (B) average $C_\alpha$-RMSD values β1, β2, and Turn region in control and Aβ –BCL system.

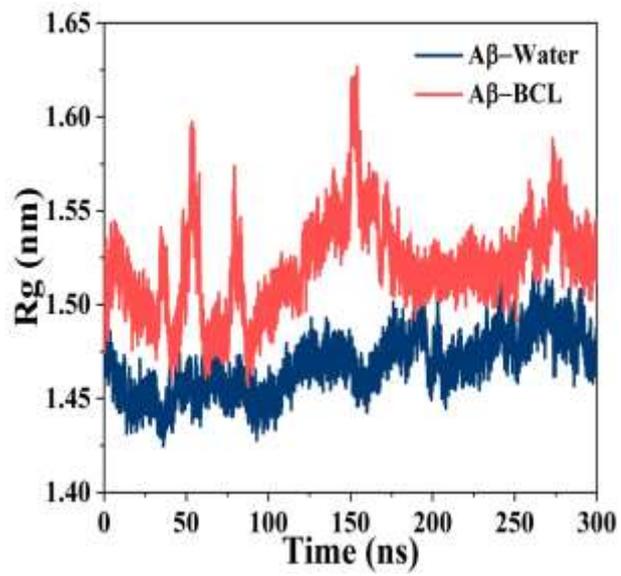 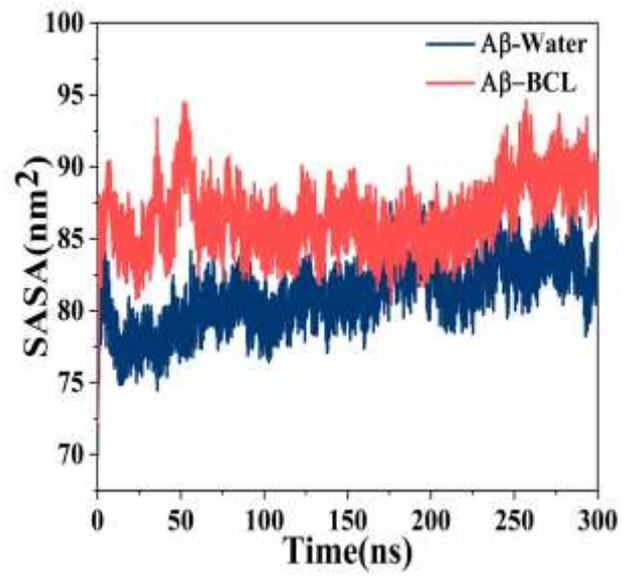

**Fig. 5: (A)** Radius of gyration (R$_g$), and the **(B)** SASA values of the protofibril in control and Aβ-BCL systems over 300 ns.

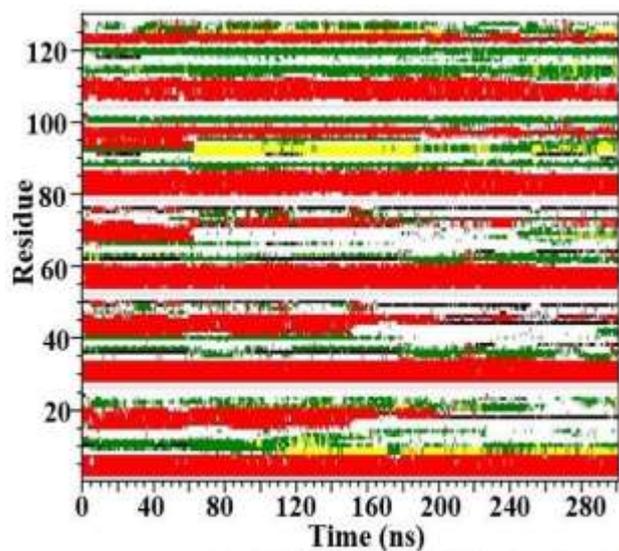 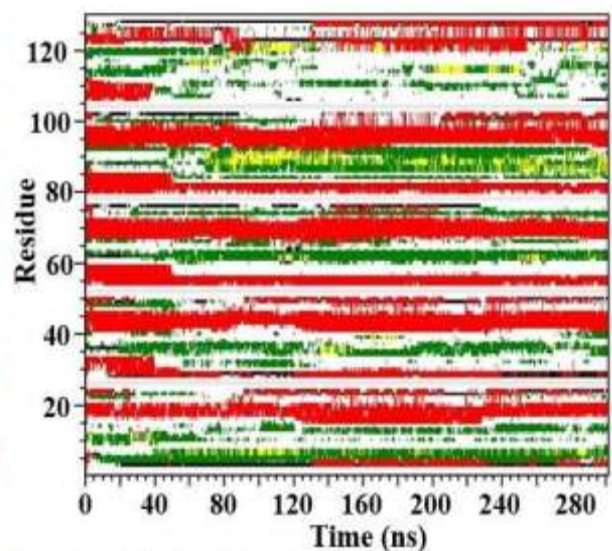

**Fig. 6: DSSP secondary structure for (A) Aβ –Water, and (B) Aβ-BCL system over 300 ns.**

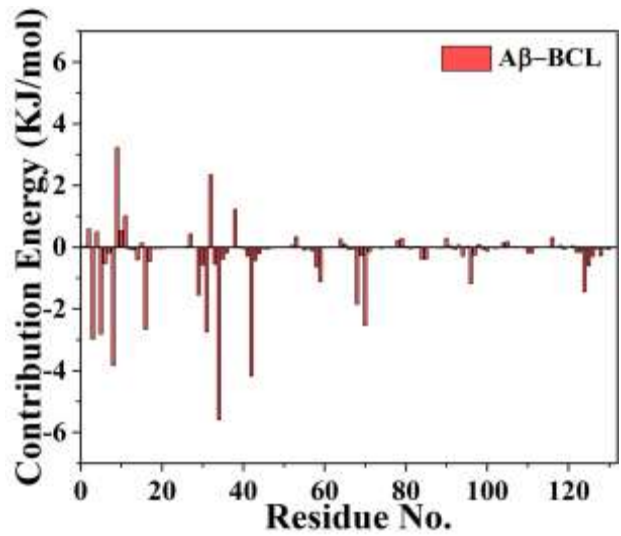

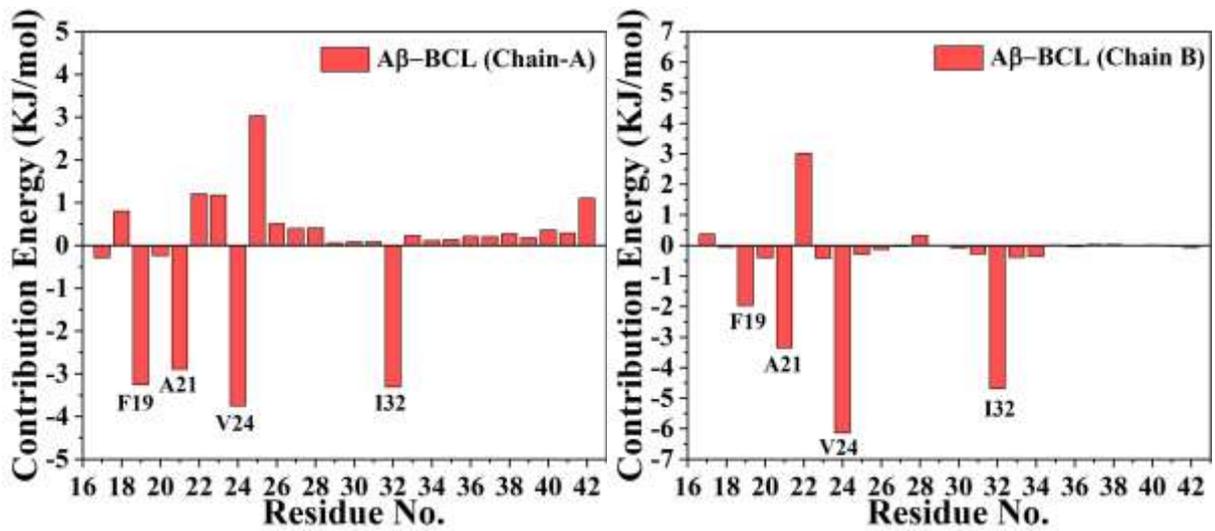

**Fig. 7: Individual residue contribution of (A) Entire fibril length, (B) Chain A, and (C) Chain B to the binding free energy of the protofibril and BCL molecule.**

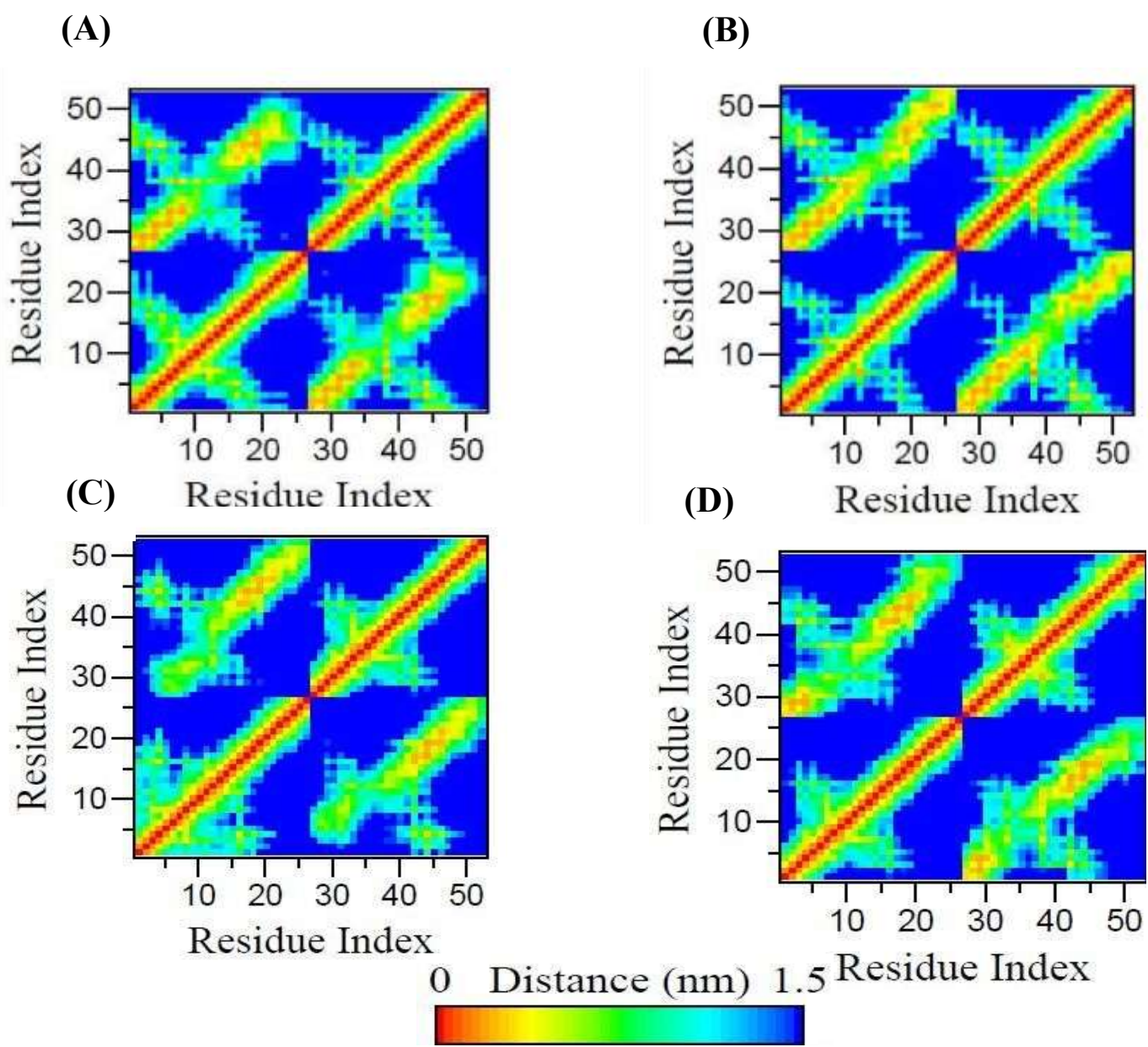

**Fig. 8: Inter-chain distance matrix for (A) Chain A-B, (B) Chain B-C of the control and (C) Chain A-B, and (D) Chain B-C of the Aβ-BCL systems.**

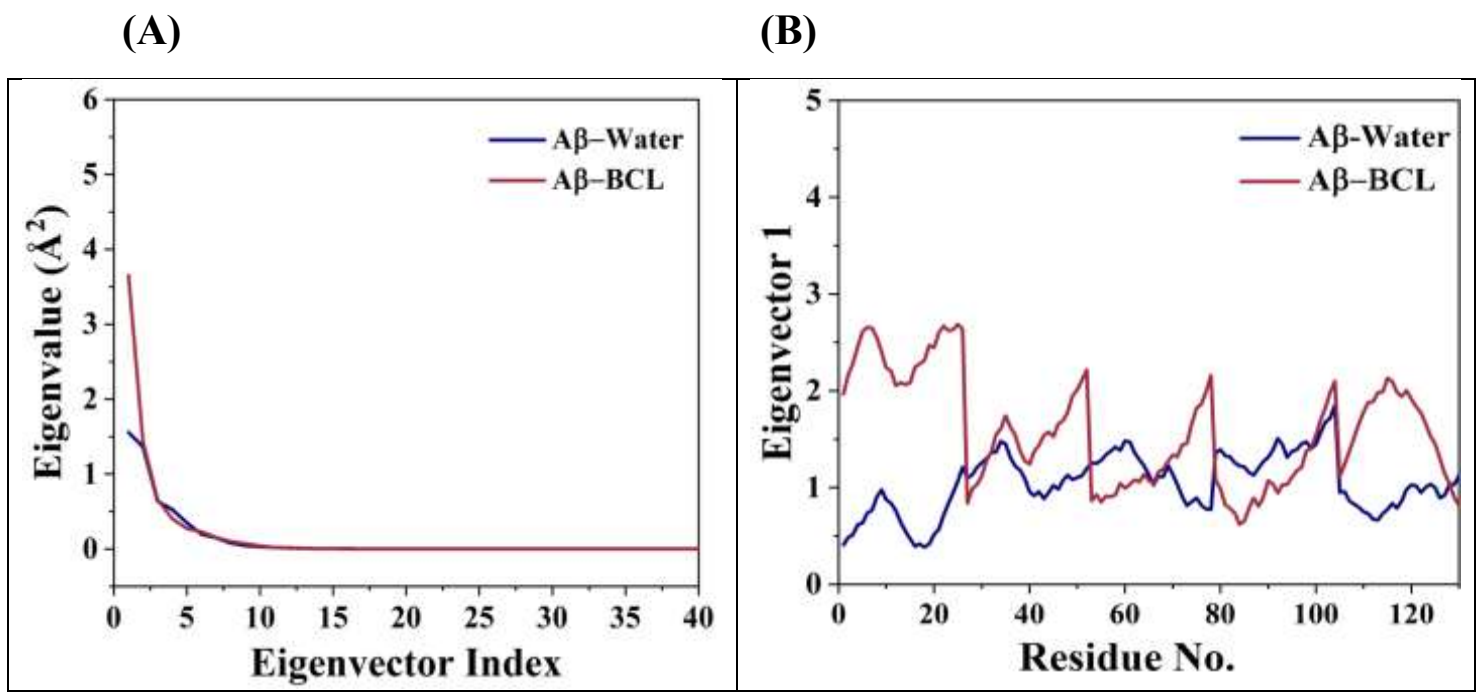

**Fig. 9: (A) The eigenvalues plotted against the corresponding eigenvector indices, and (B) Displacements of the components of the Aβ-Water and Aβ-BCL for the first eigenvector.**

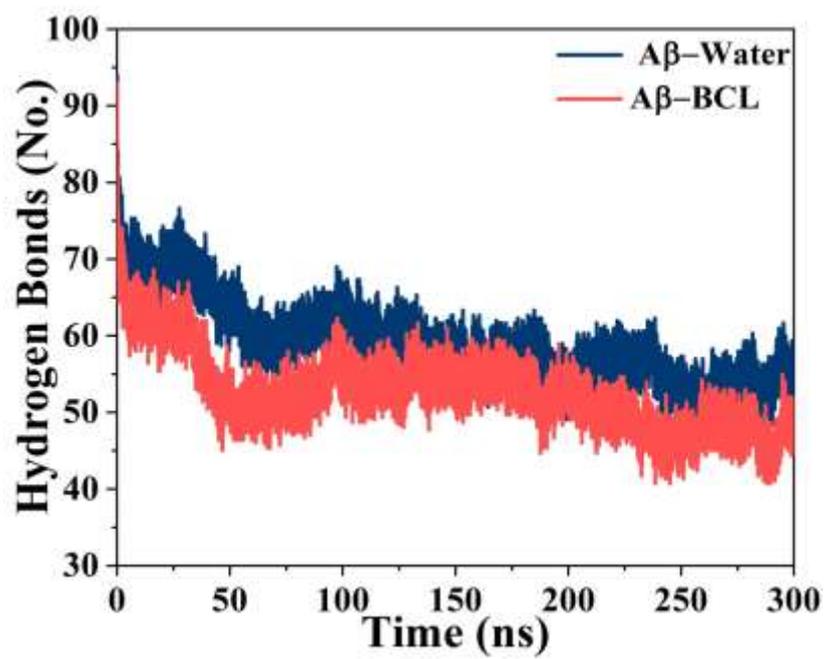

**Fig. 10: Change in the number of hydrogen bonds for Aβ-Water and Aβ-BCL systems over 300 ns simulation.**

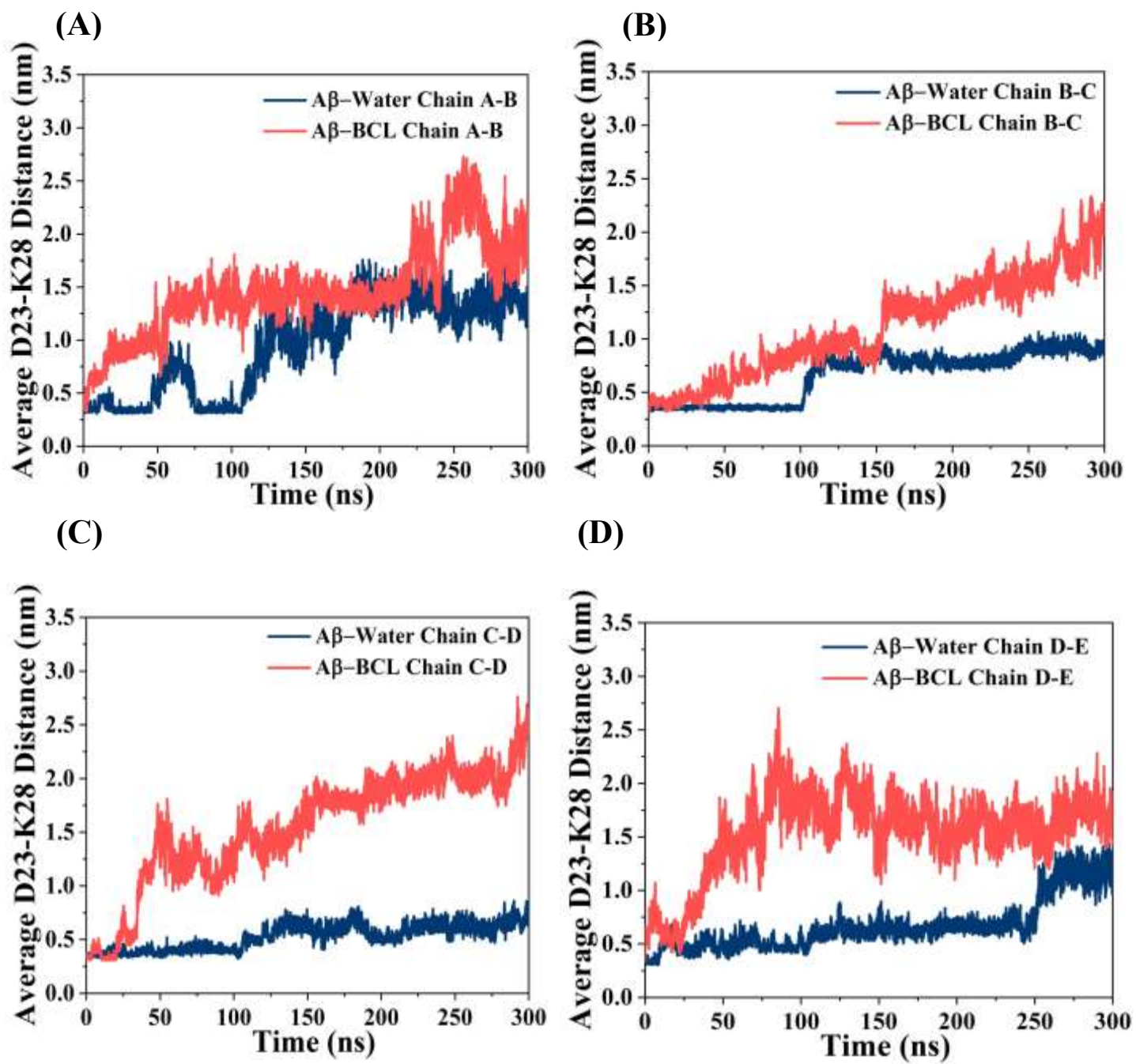

**Fig. 11:** The average inter-chain D23-K28 distances of the Aβ-Water and Aβ-BCL systems over 300 ns for (A) Chain A-B, (B) Chain B-C, (C) Chain C-D, and (D) Chain D-E.

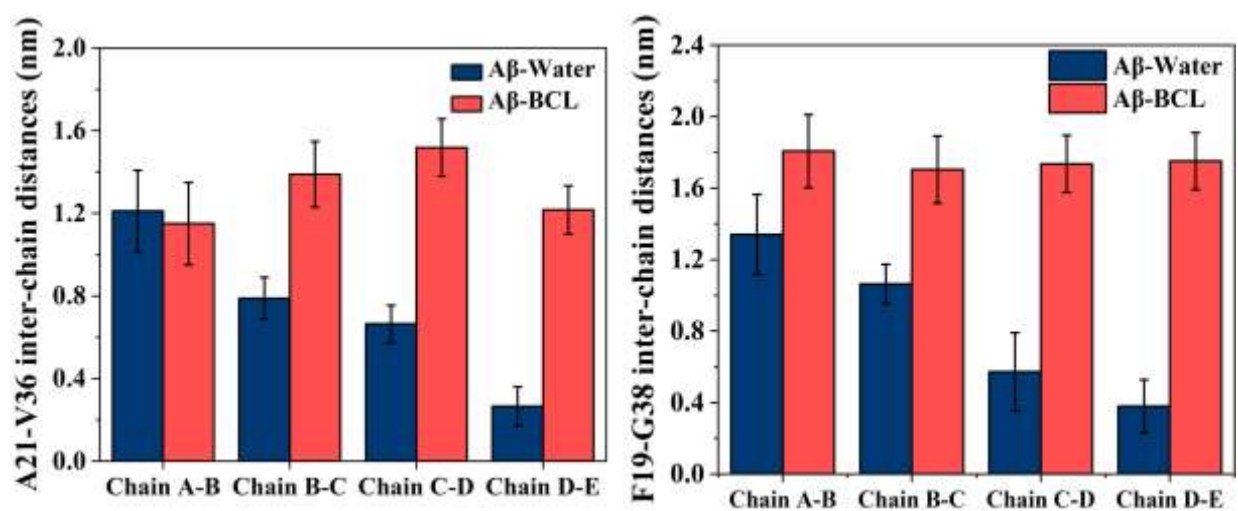

**Fig. 12: Average inter-chain distances between (A) A21-V36, and (B) F19-G38 residues.**

# Supporting Information

# Destabilization of Alzheimer's Amyloid-β Protofibrils by Baicalein: Mechanistic Insights from All-atom Molecular Dynamics Simulations

Sadika Choudhury[1] and Ashok Kumar Dasmahapatra[1,2*]

[1]Department of Chemical Engineering and [2]Center for Nanotechnology, Indian Institute of Technology Guwahati, Guwahati - 781039, Assam, India.

**Fig. S1.** Contacts made by the docked BCL molecule with the protofibril are illustrated. Residues that interact with the BCL molecule are depicted with red spikes, while hydrogen bonds are indicated by green dotted lines.

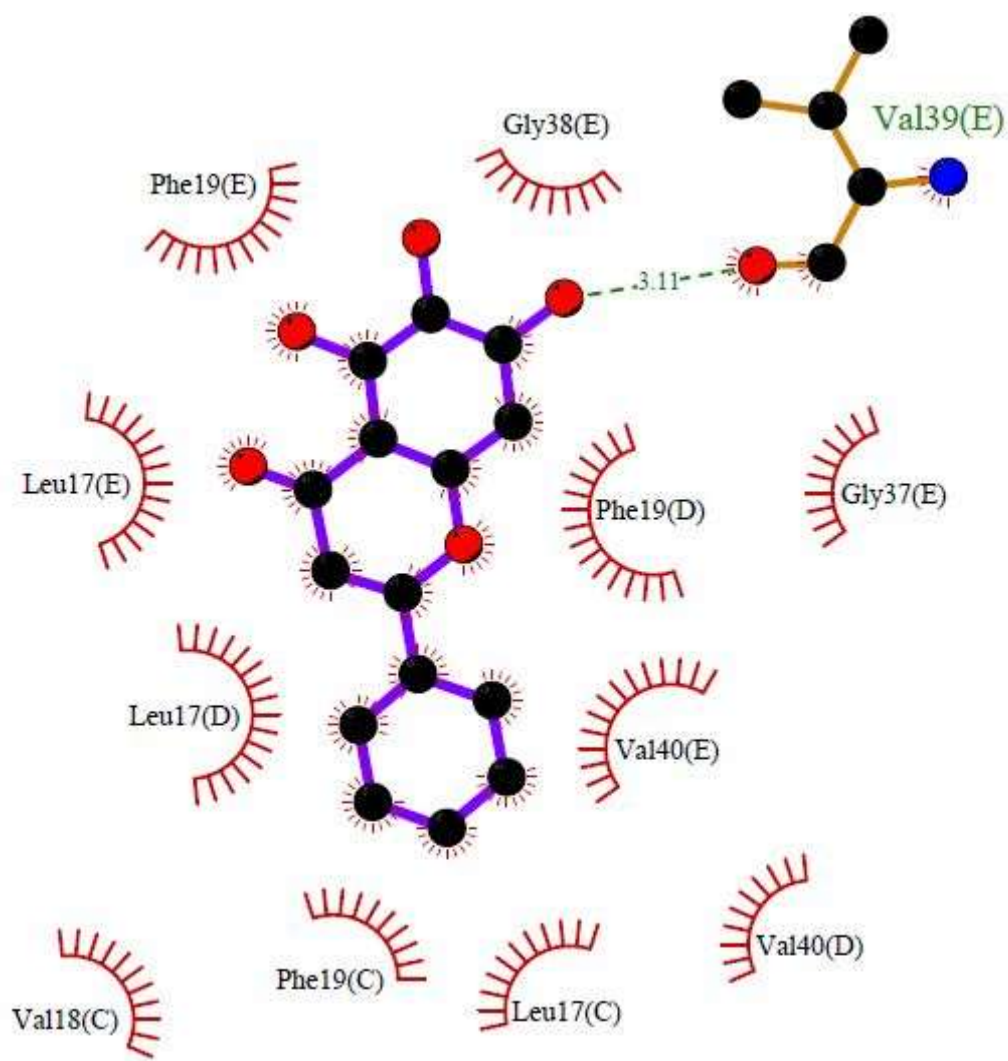

**Fig. S2.** Contacts made by the BCL molecule with the protofibril at 300ns. Residues that interact with the BCL molecule are depicted with red spikes.

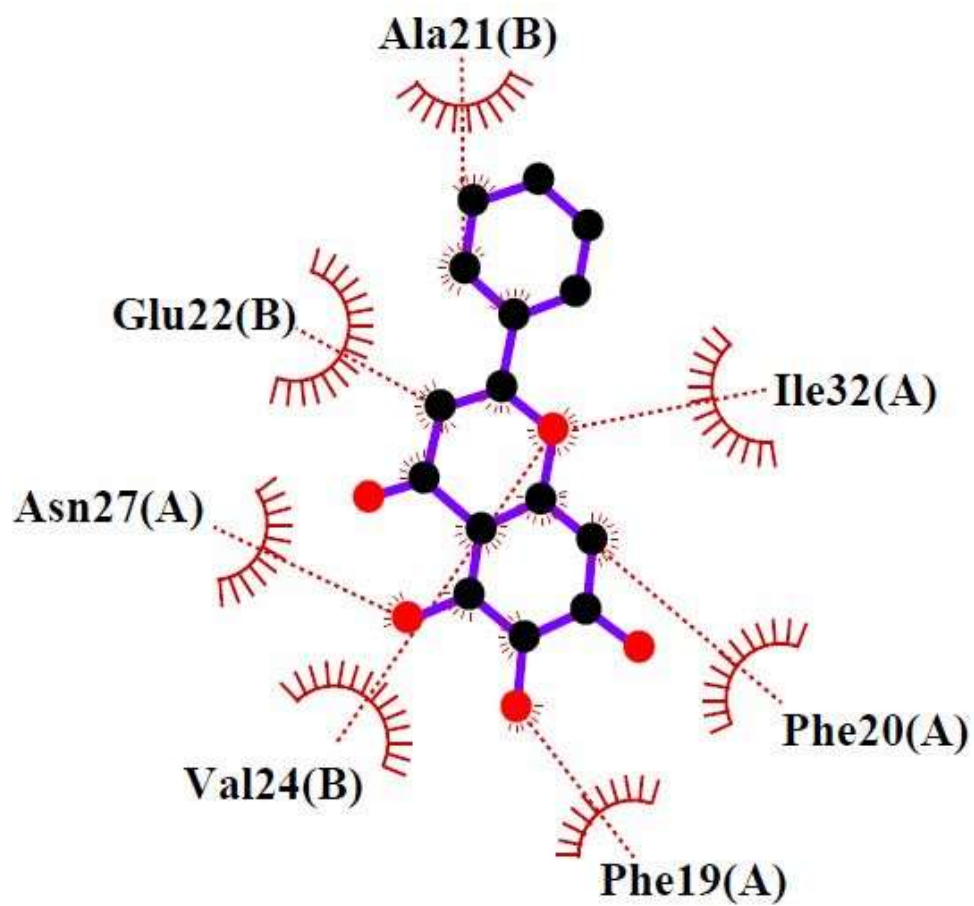

**Fig. S3. Interaction between the protofibril and BCL molecule at (A) 0ns and (B) 290ns. In Fig. S3A, we observe that the F19 residue initially establishes hydrophobic contacts with the ligand, while in chain C, the F19 residue forms a pi-stacking interaction. However, in Fig. S3B, which represents the complex at 290 ns (i.e., toward the end of the simulation), we can see that the F19 residue is primarily engaged in hydrophobic interactions, without involvement in pi stacking. This observation underscores the significant role of hydrophobic interactions in driving the destabilization process.**

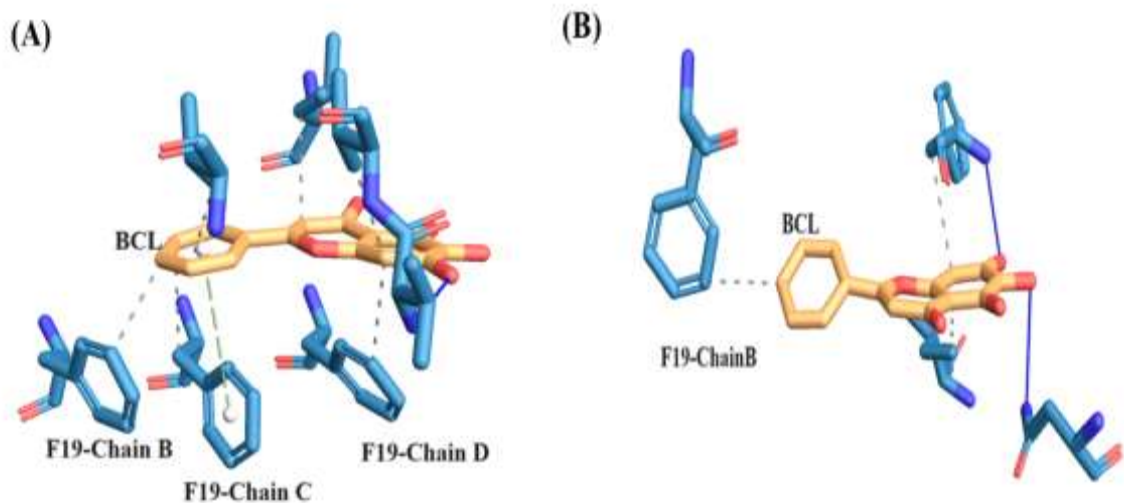

**Fig. S4.** The correlation between the predicted and experimental NMR chemical shifts in ppm for the (A) $C_\alpha$ and (B) $C_\beta$ atoms.

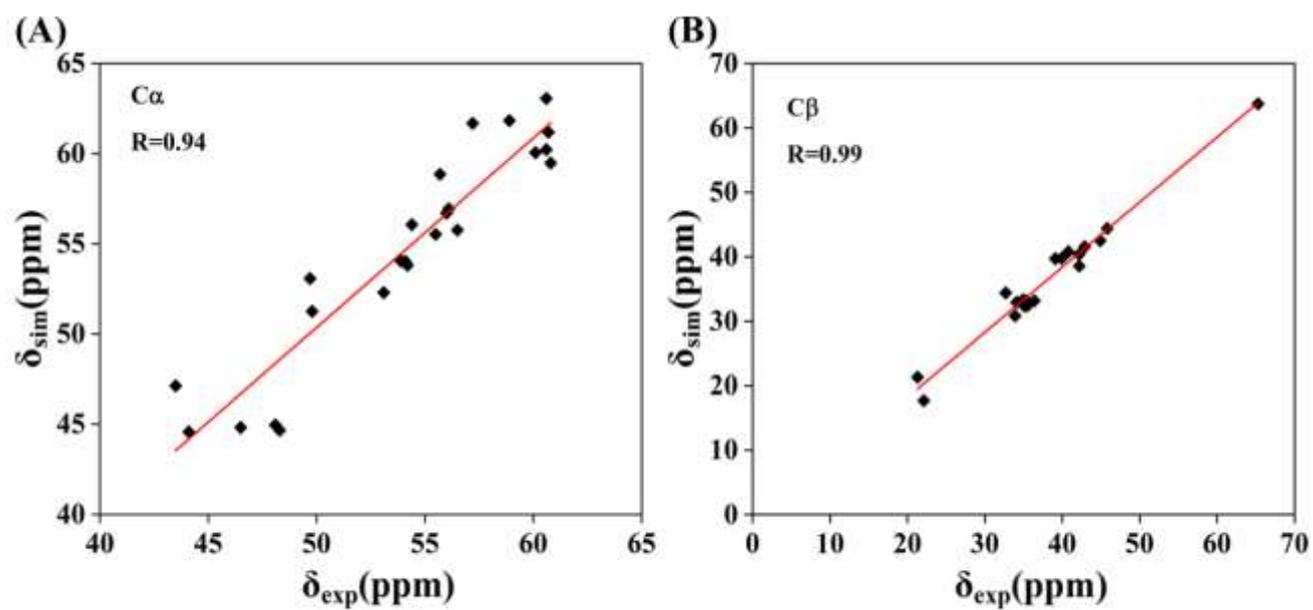

**Fig. S5.** Comparison of simulated $^3J_{HN-H\alpha}$ coupling constants for the Aβ42 in red with experimental measurements in black

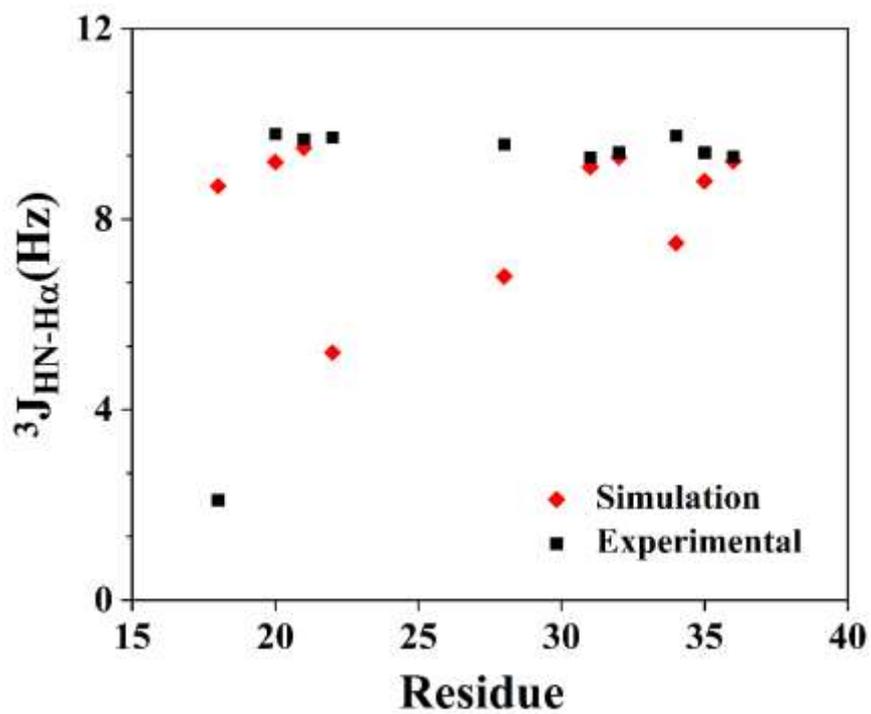

**Table S1.** Conformational snapshots of the systems at different time intervals.

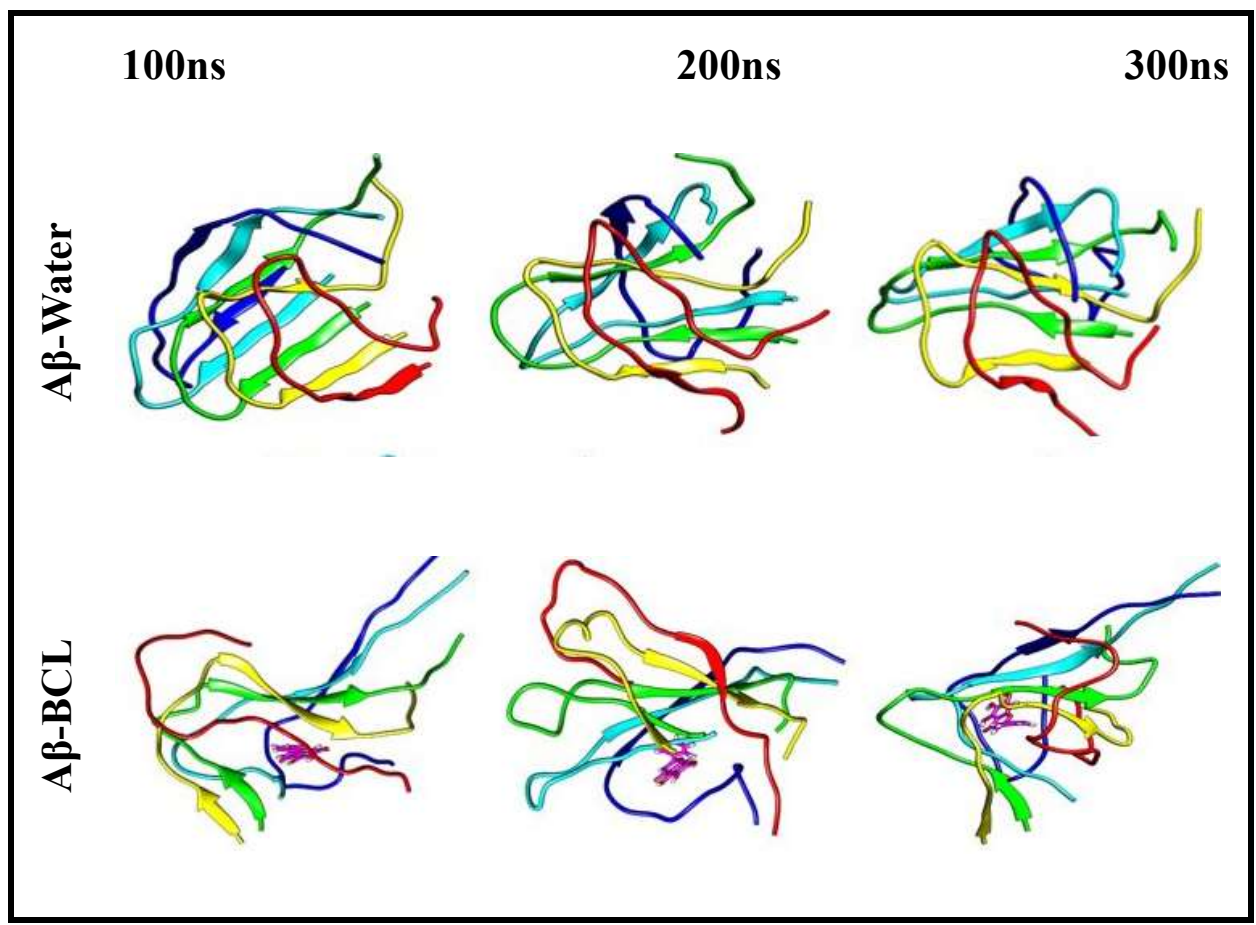

**Fig. S6.** $C_{\alpha}$-RMSD values across the 300 ns trajectory of (A) Control, the primary influence on the RMSD stemmed from the peripheral chains, with greater exposure to the solvent and (B) Aβ –BCL system. Among the three iterations, run 2 demonstrated the protofibrils with the highest RMSD values, with the most significant disturbance in the protofibril structure.

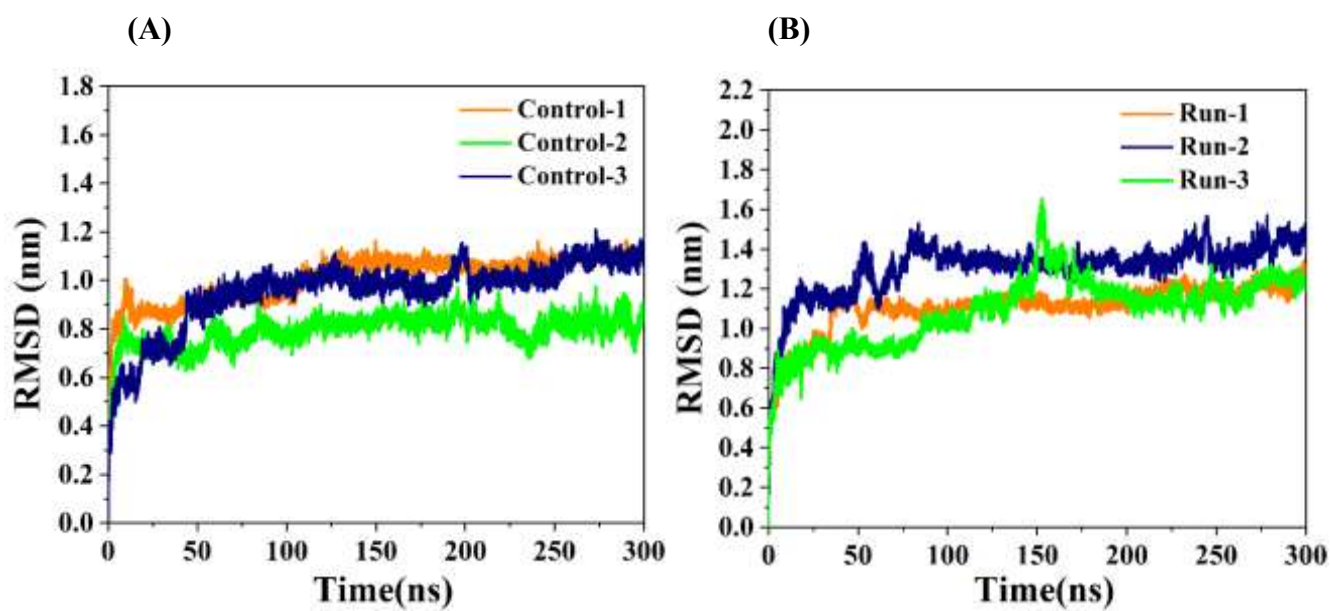

**Table S2. Radius of Gyration ($R_g$) for all iterations.**

| Systems | Radius of Gyration (nm) |
|---|---|
| **Aβ-Water** | |
| Control-1 | 1.44±0.02 |
| Control-2 | 1.46±0.03 |
| Control-3 | 1.49±0.01 |
| **Aβ-Water** | |
| Run-1 | 1.48±0.03 |
| Run-2 | 1.58±0.04 |
| Run-3 | 1.48±0.05 |

**Fig. S7. Individual residue contribution of (A) Chain C, (B) Chain D, and (C) Chain E to the binding free energy of the protofibril and BCL molecule.**

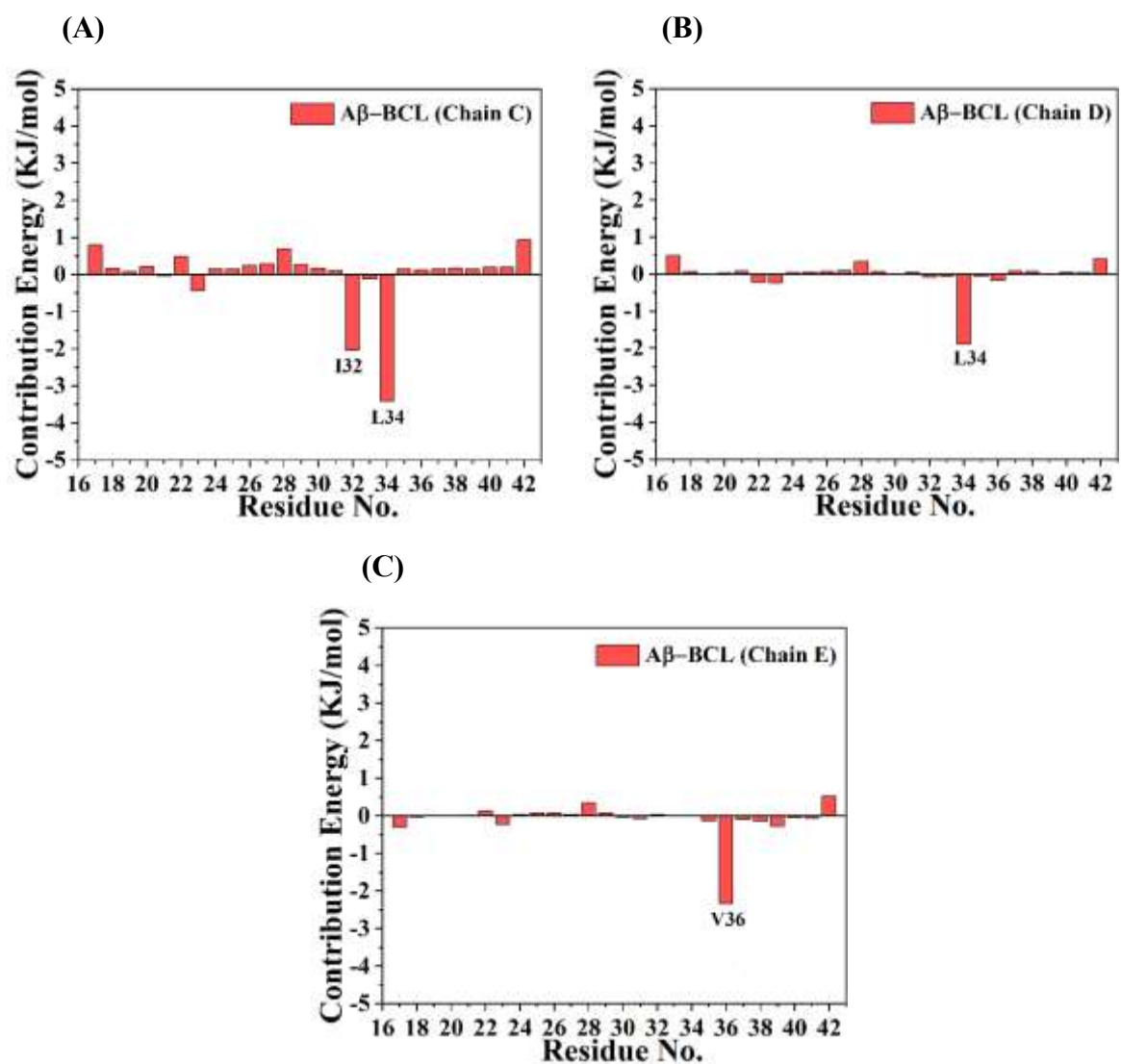

**Fig. S8.** Inter-chain distance matrix for (A) Chain C-D, (B) Chain D-E of the control and (C) Chain C-D, and (D) Chain D-E of the Aβ-BCL systems.

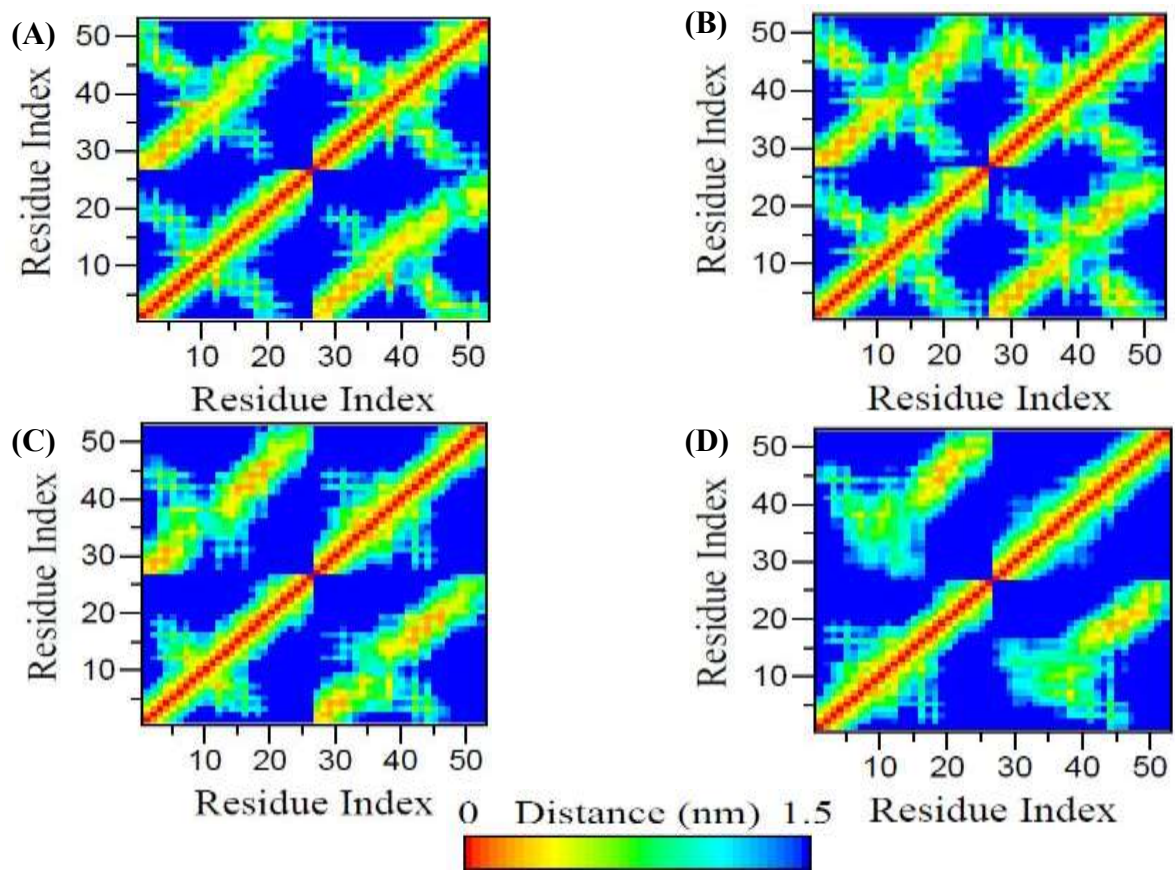

**Table S3.** The average inter-chain binding free energy (BFE) of the Aβ protofibril in the presence and absence of the BCL molecule.

| Systems | Chain A:B (KJ/mol) | Chain B:C (KJ/mol) | Chain C:D (KJ/mol) | Chain D:E (KJ/mol) |
|---|---|---|---|---|
| **Aβ-Water** | -249.88 ± 20.48 | -363.44 ± 21.81 | -325.06 ± 20.87 | -321.29 ± 31.73 |
| **Aβ-BCL** | -221.24 ± 17.48 | -219.22 ± 20.57 | -261.79 ± 26.92 | -214 ± 37.57 |

**Table S4. Average number of Inter-chain hydrogen bonds.**

| System | Chain A:B | Chain B:C | Chain C:D | Chain D:E |
|---|---|---|---|---|
| **Aβ-Water** | 11±2 | 15±1 | 11±2 | 11±2 |
| **Aβ-BCL** | 9±1 | 10±1 | 11±2 | 10±2 |